\documentclass[pra, twocolumn, showpacs, preprintnumbers, amsmath, amssymb, superscriptaddress, aps]{revtex4}
\usepackage{amssymb}
\usepackage{latexsym}
\usepackage{color}
\usepackage[tight]{subfigure}
\usepackage[normalem]{ulem}	

\usepackage{epsfig}

\usepackage{dcolumn}
\usepackage{amsmath}
\usepackage{amsfonts}
\usepackage{bm}

\newcommand{\be}{\begin{equation}}
\newcommand{\ee}{\end{equation}}
\newcommand{\bea}{\begin{eqnarray}}
\newcommand{\eea}{\end{eqnarray}}

\newcommand{\p}{\partial}

\newcommand{\ri}{i}
\newcommand{\re}{\mbox{e}}

\newcommand{\gt}{\theta}

\begin{document}
\title{Charge-Density Wave and One-dimensional Electronic Spectra in Blue Bronze: Incoherent Solitons and Spin-Charge Separation}
\author{Daixiang Mou}
\affiliation{National Laboratory for Superconductivity,
Beijing National Laboratory for Condensed Matter Physics, Institute of Physics,
Chinese Academy of Sciences, Beijing 100190, China}
\author{R. M. Konik}
\affiliation{CMPMS Dept., Brookhaven National Laboratory, Upton, NY 11973-5000, USA$^2$}
\author{A. M. Tsvelik$^{*}$}
\affiliation{CMPMS Dept., Brookhaven National Laboratory, Upton, NY 11973-5000, USA$^2$}
\author{I. Zaliznyak}
\affiliation{CMPMS Dept., Brookhaven National Laboratory, Upton, NY 11973-5000, USA$^2$}
\author{Xingjiang Zhou$^{*}$}
\affiliation{National Laboratory for Superconductivity,
Beijing National Laboratory for Condensed Matter Physics, Institute of Physics,
Chinese Academy of Sciences, Beijing 100190, China}
\date{\today }
\begin{abstract}
We present new high resolution angle resolved photoemission (ARPES) data for K$_{0.3}$MoO$_3$ (blue bronze) and propose a novel theoretical description of these results.
The observed Fermi surface, with two quasi-one-dimensional sheets, is consistent with a ladder material with a weak inter-ladder coupling. 
Hence, we base our description on spectral properties of one-dimensional ladders.
The  marked broadening of the ARPES lineshape, a significant fraction of an eV, is interpreted in terms of
spin-charge separation.
A high energy feature, which is revealed for the first time in the spectra near the Fermi momentum thanks to improved energy resolution, is seen as a signature of a higher energy bound state of soliton excitations on a ladder.
\end{abstract}


\maketitle

Systems of two coupled chains (called ladders) can be viewed as a first step in crossing over from one dimension (1D), with  its exotic physics of Luttinger liquid and spin-charge separation, to higher dimensions. In spite of being unusual and seemingly enigmatic, this 1D physics is now  well understood, thanks to remarkable progress in applying field theoretic methods in condensed matter \cite{tsvelik_book}. How this physics transforms in the course of crossover to higher dimension is why ladders have been intensely studied. While theoretical progress on this problem has been considerable, with work in predicting rich excitation spectra, dynamical generation of spectral gaps, existence of preformed pairs \cite{fabrizio,fisher,schulz,Khev,konik2,varma,controzzi,lee,furusaki,wu,noack,jeckelmann,weihong,poilblanc,2leg}, the progress on the experimental side has been slower both due to limited numbers of materials with a suitable structure and limited experimental accuracy.

The molybdenum blue bronzes, A$_{0.3}$MoO$_3$ are among the most interesting and intensely studied examples of quasi-1D conductors \cite{Greenblatt,Monceau2012}. They feature MoO$_6$ octahedra forming an array of weakly coupled pairs of conducting ladders \cite{Graham}. According to band structure calculations  the chemical potential is crossed by two slightly warped bands with 3/4-filling \cite{Canadell,Mozos}. These are bonding (B) and anti-bonding (AB) bands arising from the electron hopping between the two legs of a given ladder, with Fermi wave vectors $K_{FB}$ and $K_{FAB}$, respectively. The resulting Fermi surface nesting suggests a charge density wave (CDW) formation with a wave vector $K_{FB}+K_{FAB}$ along the chains.

In K$_{0.3}$MoO$_3$ a 3D phase transition into a charge-ordered insulating phase takes place at a temperature T$_{CDW} \approx 180$K \cite{Johnston,Kwok} even though the magnetic susceptibility starts to decrease well above the transition, experiencing a one-third drop in the interval between 700K and 180K \cite{Johnston}. 
The low-energy lattice responses measured in neutron and X-ray experiments also show precursor effects up to at least $\sim 2$T$_{CDW}$ \cite{Monceau2012,Pouget,Fleming1985,Sato}. Previous angle-resolved photoemission (ARPES) measurements, while confirming the band structure predictions, also found significant breadth of the electronic spectra, extending to photoelectron energies $\sim $ eV \cite{Fedorov,Perfetti,Ando}. By virtue of the lattice involvement in the CDW formation, this surprising incoherence was assigned to electron-phonon interactions and small polarons, in spite of the marked mismatch in the energy scales.

Here we report the results of new high resolution ARPES measurements of K$_{0.3}$MoO$_3$, which we analyze based on the electronic spectral properties of ladders. With much improved instrumental resolution, we focus upon examining the fine structure of the measured photoemission spectra. We resolve for the first time two broad peaks in regions of the Brillouin zone near the Fermi vector, dispersing through energies a significant fraction of an eV. We interpret these features as ladder excitations originating from the same interaction that underpins the 3D charge order existing in this material at low temperatures. The breadth of these peaks arises from the presence of gapless charge excitations (holons) and gapful spin solitons (spinons) with markedly different velocities.  We obtain a good description of the measured spectra with two SU(2) Thirring models describing the spectral properties of a ladder, thus assuming that the physics is electron-driven and the electron-lattice interaction plays only a secondary role.

The K$_{0.3}$MoO$_3$ single crystals were grown by an electrolytic reduction method \cite{Li}. High resolution angle-resolved photoemission
measurements were carried out with a Scienta R4000 electron energy analyzer \cite{Liu}. For the band structure measurements and Fermi surface mapping (Fig. 1), we used a helium discharge lamp as the light source with a photon energy of $h\nu$ =21.218 eV. The overall energy resolution  was set at 10 meV and the angular resolution was 0.3$^\circ$, corresponding to a momentum resolution of 0.0091 \AA$^{-1}$ at the photon energy of 21.218 eV. For the high-precision ARPES measurements (Fig. 2 and Fig. 3), a vacuum ultra-violet (VUV) laser  with a photon energy $h\nu$=6.994 eV was used as a light source. The energy resolution in this case was 1 meV. The angular resolution is 0.3$^\circ$, corresponding to a momentum resolution of 0.004 A$^{-1}$ at the photon energy of 6.994 eV. The Fermi level is referenced from measuring a clean polycrystalline gold that is electrically connected to the sample. The crystal was cleaved {\it in situ} and measured in vacuum with a base pressure $\lesssim 5 \times$10$^{-11}$ Torr.

\begin{figure}[htp]
\centerline{\includegraphics[angle = 0,
width=1\columnwidth]{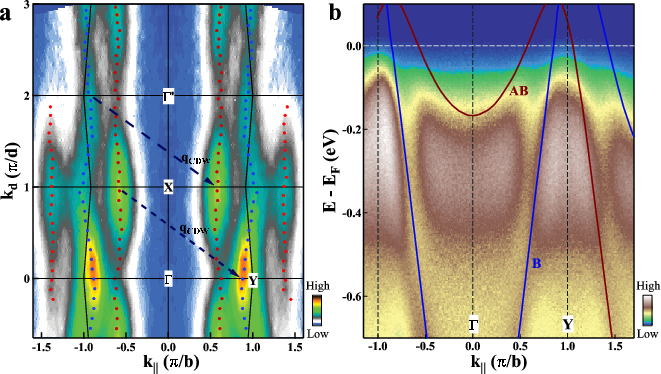}}
\caption{
a) The measured Fermi surface of K$_{0.3}$MoO$_3$. Data are symmetrized with respect to the k$_{||}$=0 line. Blue and red dots show the extracted Fermi momentum along the Fermi surface sheets. The CDW vectors
are marked by the dashed arrows. b) The measured band structure along $\Gamma$-Y (k$_d$ = 0) showing two bands, the bonding (B) and the anti-bonding (AB) ones, crossing the Fermi level. The solid lines mark bands predicted from band structure calculations \cite{Mozos}.
}
\end{figure}
Fig. 1a shows the bare Fermi surface (FS) map measured at 80 K.
The FS is open and consists of weakly warped sheets (blue and red dots in Fig. 1a).  This warping arises from a weak interladder
hopping, $t_{\perp} \approx 100$ meV (to be compared with the hopping between the legs of a given ladder, $t_{\rm rung} \approx 500$ meV).
Fig 1b shows the band structure measured along the $\Gamma$-Y cut at 80 K. At energies less than 1eV we clearly distinguish two bands (B and AB). The non-interacting bands cross the chemical potential at 0.97 $\pi$/b, 0.6 $\pi$/b, -0.55 $\pi$/b and -0.96 $\pi$/b, consistent with the band structure calculations in Ref. \onlinecite{Mozos} and previous ARPES measurements \cite{Fedorov,Perfetti,Ando}. The Fermi surface nesting vector is thus estimated to be $\approx$ 1.52 $\pi$/b, also in good agreement with the measured CDW vector \cite{Pouget,Fleming1985,Sato}.

\begin{figure}[b]
\centering
\centerline{\includegraphics[angle = 0,width=.9\columnwidth]{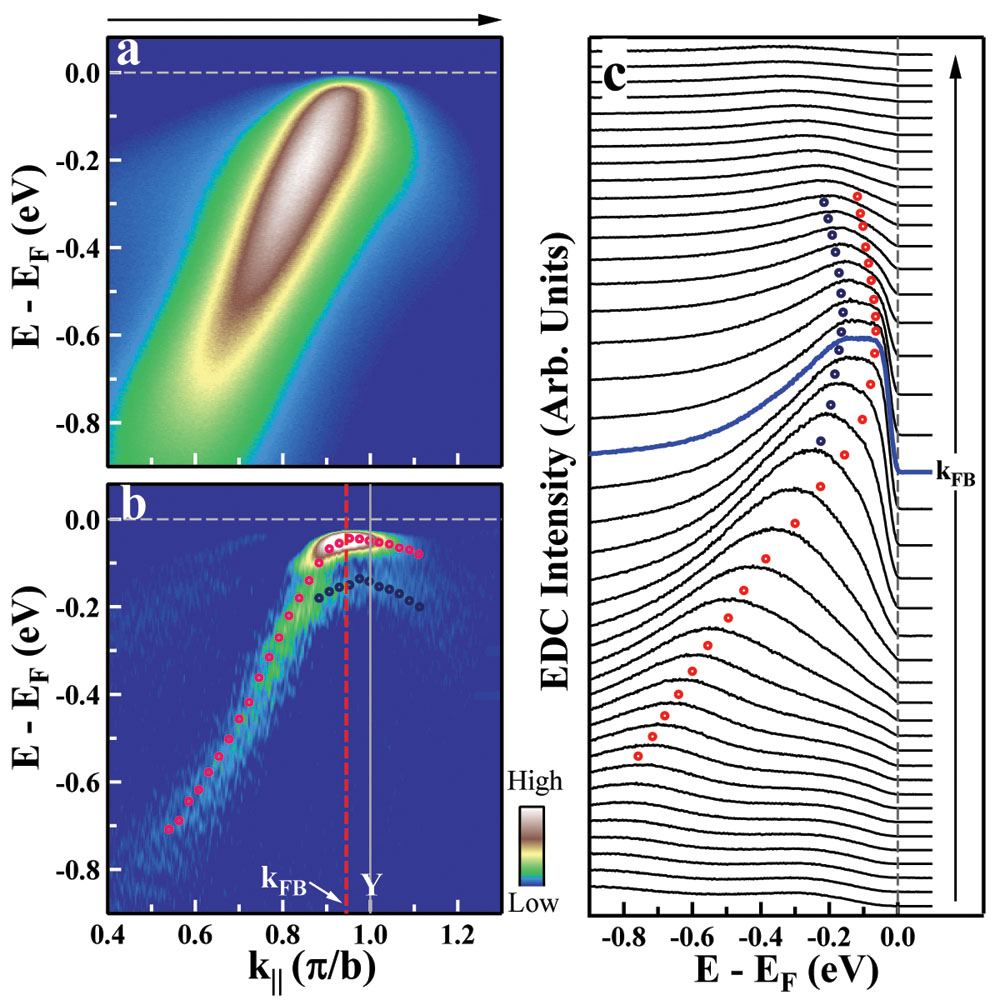}}
\caption{a) The band structure along $\Gamma$-Y momentum cut measured with 1 meV energy resolution. Due to ARPES matrix element effects, the AB band intensity is suppressed. 
b) Second derivative with respect to energy of the data in (a). Two structures around k$_{FB}$, corresponding to incoherent soliton and bound state (see text), are clearly seen. 
c) The corresponding photoemission spectra (EDCs). 
Red (blue) circles mark in b) and c) the dispersion of the soliton and the bound state, respectively, both for the B and the weak AB band. 
}
\end{figure}

\begin{figure}
\centering
\centerline{\includegraphics[angle = 0, width=.9\columnwidth]{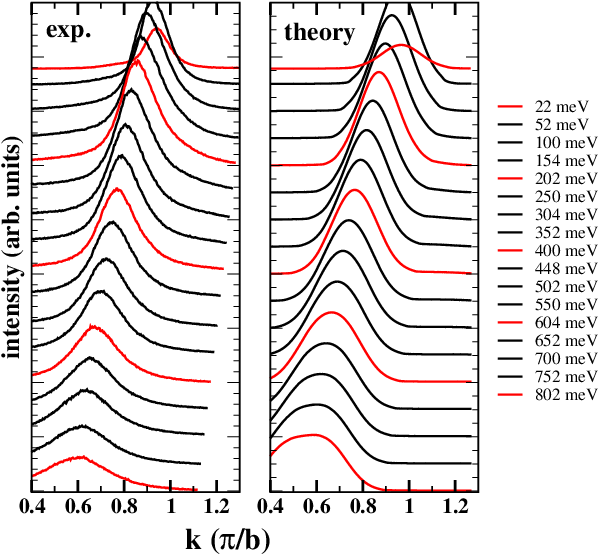}}
\caption{The MDCs for different binding energies obtained from the data in Fig. 2 (left) and the theoretical prediction for these MDCs (see the text)  and accounting for the surface inhomogeneity of the Fermi wave vector. }
\label{fig:MDC}
\end{figure}

In Fig. 2 we present high resolution laser ARPES data for the band structure along the ladder. 
Due to the effects of matrix elements, the B band is more clearly visible than the AB band (Fig. 2a). Its electronic structure exhibits two features, seen as excitation branches near $K_{FB}$,  clearly distinguished in the second derivative image (Fig. 2b). The lower binding-energy branch was observed in earlier  measurements \cite{Fedorov,Perfetti,Ando}, while the higher energy branch has not hitherto been resolved. The lower branch, starting at $\approx 60$ meV below $E_F$, reaches binding energy $\approx 0.8$ eV as $k_{||}$ varies from $\pi/b$ to 0.4 $\pi/b$ (red circles in Fig. 2b). The second branch with a lower spectral weight starts at $E - E_F \approx -150$ meV, remains distinct until roughly $\approx - 200$ meV, and then merges with the first branch (blue circles in Fig. 2b). The two branches gradually become indistinguishable when the  temperature increases so that at 240 K only one broad hump is observed (see supplementary material \cite{supp}). We argue that the lower branch is due to the smeared contribution of the holon and spinon to the ARPES signal while the upper branch is due to a spinon plus a coherent excitation formed as a bound state of two spinons.

A further feature not previously seen  is the left-right asymmetry of the low energy branch: the slope of the band on the right side of the Fermi momentum ($k_{||} > K_{FB}$) $v_R\approx$ 1.15 eV$\AA$ is smaller than that on the left side ($k_{||} < K_{FB}$): $v_L \approx$ 5.33 eV$\AA$. These roughly correspond to the expected velocities of the B and AB bands \cite{Mozos}. We  argue that both the presence of two excitation branches and the asymmetry of their dispersion can be explained in terms of 1D intra-ladder interactions.

The previous attempts to explain the broad ARPES spectrum have been based on small polaron theory \cite{Perfetti,Roscha}. However, the shear size of the gaps and the energy dependence of the incoherence peaks' widths, running to a significant fraction of an eV, suggests that these features have an electronic origin. Here we present non-perturbative calculations  for the ARPES response. Before presenting theoretical details, we present in Fig. \ref{fig:MDC} the momentum distribution curves (MDCs) obtained by cutting the measured ARPES spectra at a number of different binding energies and our theoretical predictions for these MDCs.  We see an impressive fit between the two.

We adopt an approach where we consider an interacting ladder as the basis for calculations, treating it using powerful 1D non-perturbative techniques, and then considering interladder interactions as a perturbation \cite{Essler2002,Essler2005,Konik}.
In order to account for the experimental observations, we devise a low energy theory that (i) maximizes the fluctuations leading to the observed incoherency of the ARPES lineshapes and (ii) gives rise to a zero temperature CDW instability at $k_{FB}+k_{FAB}$ which, in the presence of 3D coupling, is shifted to finite temperatures. Since the CDW instability occurs at the wave vector $k_{FB} + k_{FAB}$, both bands are involved in its creation. However, as is clearly seen on Fig. 1b, these bands have very different Fermi velocities $v_{FB}$ and $v_{FAB}$ - a feature that is not present in standard theories of ladders. This velocity difference precludes the emergence of an SO(6) symmetry, which is one of the most interesting predictions of the low energy theory of ladders \cite{fisher,schulz}. Nevertheless, we will argue that a remnant of this symmetry remains in the form of an electron-hole bound state.

\begin{figure}[t]
\centerline{\includegraphics[angle=0,width=.85\columnwidth]{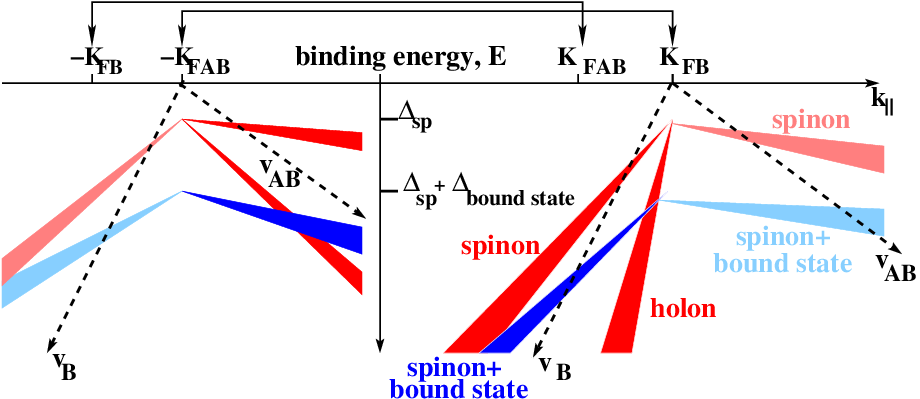}}
\caption{Schematics of interaction between pairs of Fermi points, ($K_{FB},-K_{FAB}$) and ($K_{FAB},-K_{FB}$).  For the first pair we show the asymmetry in the dispersion on either side of these two Fermi points. The color bars locate the contributions in $(k,E)$ of
the various excitations (holon (red), spinon (red), spinon + bound state (blue)) to the ARPES signal, with the intensity of the color corresponding to the intensity of the ARPES signal.
}
\end{figure}

Since the spectral gaps are much smaller than the bandwidth, it is reasonable to linearize the dispersion close to the Fermi points by introducing slowly varying right- and left-moving components of the fermionic fields
\begin{equation}
\psi_{j\sigma} = \re^{-\ri k_{Fj}x}\psi_{Rj\sigma} + \re^{\ri k_{Fj}x}\psi_{Lj\sigma}, ~~ j=B,AB.
\end{equation}
Because of the different Fermi velocities in the bonding and anti-bonding bands, we expect the
theory will not have a symmetry higher than U(1)$\times$SU(2).  Moreover, we
expect the strongest interaction to occur between
the bonding and anti-bonding bands, foreshadowing the rise of the observed 3D CDW order.  To this end we consider
a model where the fermions at $K_{FB}$ interact with the fermions at $-K_{FAB}$ and similarly those
at $-K_{FB}$ interact with fermions at $K_{FAB}$.  This leads us to describe the system as two decoupled 1D theories,
$H=H_1+H_2$:
\begin{eqnarray}\label{GNmodel}
H_1 \! &=& \!\! -i\int\!\! dx \big(v_{FB}\psi^\dagger_{Rp\sigma}\p_x \psi_{Rp\sigma}\! -\! v_{FAB}\psi^\dagger_{LAB\sigma}\p_x \psi_{LAB\sigma}\big) \cr
&+& \sum_{l,l'=R,L}g_{ll'0}\rho_{lB}\rho_{l'AB} + g_{ll'} J^a_{lB}J^a_{l'AB};\cr
\rho_{lp} &=& \psi^\dagger_{lp\sigma}\psi_{lp\sigma}, ~~ J^a_{lp} = \psi^\dagger_{lp\sigma}\tau^a_{\sigma\sigma'}\psi_{lp\sigma'};
\end{eqnarray}
with $H_2 = H_1 (AB\leftrightarrow B)$ and
where $\tau^a$ are Pauli matrices.  The coupling constants $g_{ll}$ and $g_{0ll}$ renormalize the spin
and charge velocities (see Ref. \cite{supp}).  On the other hand the couplings $g_{LR},g_{0LR}$ determine
the presence of spin and charge gaps.  $g_{LR}$ is equal to the $K_{FB}+K_{FAB}$
Fourier component of the renormalized
Coulomb interaction. We take it to be positive, corresponding to an attractive
interaction that is necessary for CDW formation.  It leads to a
spinon mode with gap $\Delta \sim 0.1 \Delta_{1/2}$ \cite{esskon}
where $\Delta_{1/2}\sim W(g_{LR}/v_{Av})^{1/2}\exp(-\pi v_{Av}/g_{LR})$ \cite{review}
(with $v_{Av} = (v_{FB} + v_{FAB})/2$) is the putative gap at 1/2 filling. We assume $g_{0LR}$ is negative, consistent
with gapless holons.  The magnitudes of these various couplings mark blue bronze
as moderately interacting \cite{supp}.

\begin{figure}
\subfigure{\includegraphics[width=0.32\textwidth]{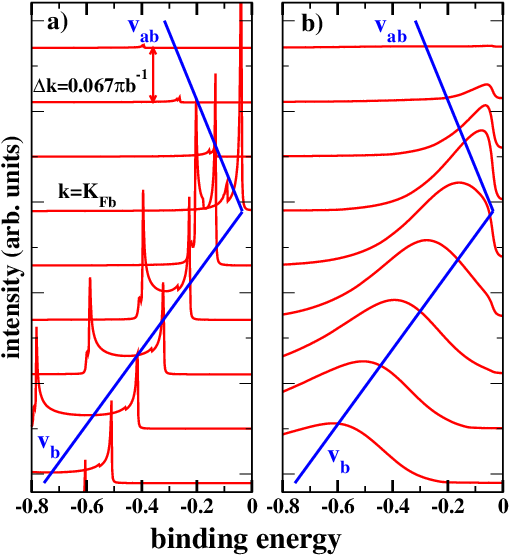}}%
\subfigure{\includegraphics[width=0.18\textwidth,height=0.35\textwidth]{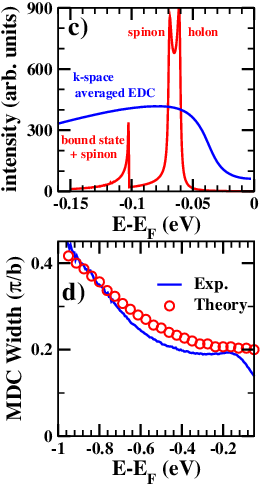}}%
\caption{a) and b): EDCs for the B  band for k-vectors ranging from $k=0.64\pi b^{-1}$ to
$k=1.17\pi b^{-1}$: a) before averaging
for differential surface doping; b) after averaging with a Gaussian of width $0.07\pi b^{-1}$.
c) Different contributions to EDC at $k=0.94\pi b^{-1}$: holon, spinon, and spinon and bound state. d) Comparison of MDC
width between theory and experiment.}
\end{figure}

The spectral function corresponding to this theory can be obtained from the one for the case of equal Fermi
velocities \cite{Essler2002} by a Galilean transformation:
$ t'= t, ~~ x' = t\Delta v+x, ~~ \Delta v = (v_{FB} - v_{FAB})$.
The corresponding retarded Green function is 
\begin{eqnarray}
G_{RB}(\omega,k) \!\!&=&\!\! \langle \psi_{Rb\sigma}\psi^\dagger_{Rb\sigma}\rangle(\omega,k) \!=\! \frac{2Z}{1+\alpha}F(\omega-k\Delta v/2,k);\cr
F(\omega, k) \!&=&\!
\frac{\tilde\omega+v_c k}{\Delta^2+v^2_ck^2-\tilde\omega^2}\Big[\big(\Delta+\sqrt{\Delta^2+v_c^2k^2-\tilde\omega^2})^2\cr
&& \hskip .5in -\frac{1-\alpha}{1+\alpha}(\tilde\omega+v_c k)^2\Big]^{-1/2},
\label{spectral}
\end{eqnarray}
where $Z \sim 1$, $v_{c,s,\pm}=v_{c,s}\pm \Delta v/2$, $\alpha =v_s/v_c$, and $\tilde\omega = \omega + i\delta$.
We can use the measured $v_L$ and
$v_R$, the velocities to the left and right of $k=K_{FB}$, to infer $v_s$ and $v_c$ (see \cite{supp}).

The most notable prediction  is that the spectral features on different sides of any Fermi point
disperse with different velocities.  This is illustrated schematically in
Fig. 4 for the pair of Fermi points $K_{FB}$ and $-K_{FAB}$. We also see the observed asymmetry on the spectrum in
Fig. 5 where we plot the expected line shapes for a set of EDCs for different k's in the vicinity of $K_{FB}$:
for $k<K_{FB}$ the spectral weight disperses with velocity $v_{FB}$ while for $k>K_{FB}$ the spectral weight, while
much reduced, disperses with velocity $v_{FAB}$.

We have so far ignored interactions between fermions living at $K_{FB},-K_{FAB}$ and $-K_{FB},K_{FAB}$.
However, such interactions are present even though they are weaker than those responsible for the 3D CDW order. In a ladder system
with equal velocities in the bonding and anti-bonding bands, the theory governing the low energy response has a dynamically generated SO(6) symmetry, and so predicts that there will be a bound state of the spin degrees of freedom of the electrons \cite{schulz,esskon}. This bound state leads to a feature at an energy $\omega = (1+\sqrt{2})\Delta$ for $k=K_{FB},K_{FAB}$, above the single particle gap, $\Delta$, in the spectral function. While in a theory with broken SO(6) symmetry (i.e., $v_B\neq v_{AB}$) we cannot predict where exactly this threshold will occur, we expect that this bound state will survive symmetry breaking, as a consequence of the bound state having significant binding energy. We believe that the contribution to the spectral function coming from this bound state is responsible for the band of intensity at higher binding energy in Fig. 2b (the blue dots). In Fig. 4 we schematically illustrate all of the contributions, holon, spinon, and spinon plus bound state, to the spectral function.

In Fig. 5a and 5c, we see how spin-charge separation leads to broadened features in the ARPES lineshape giving rise to two singularities in the spectral weight at an upper and a lower energy that disperse with different velocities (the spin and charge velocities) as $k$ moves away from the Fermi point.  We also see how the bound state plus spinon makes a distinct contribution to the spectral function.  (Our estimate of the contribution here made by the bound state is based on the symmetric SO(6) theory - see \cite{supp}). At $k$ near $K_{FB}$ this contribution appears at energies higher than the spinon and holon edges (Fig. 5c).  But as $k$ moves away from $K_{FB}$ this second contribution merges into the initially lower holon edge (Fig. 5a). This is precisely the behavior we observe in Fig. 2a.

However, different spin and charge velocities alone cannot explain the broad linewidths observed. Fig. 5d shows that the experimental widths of the momentum distribution curve (MDCs) saturate to a value of about $0.18\pi b^{-1}$ at low binding energies. 
We believe that additional broadening of ARPES spectra arises from charge inhomogeneities present on the surface of K$_{0.3}$MoO$_3$, which is a poor conductor. They generate a nearly rigid shift of the surface bands \cite{Monceau2012} and will lead to ladders in different regions on the surface of the sample to be at different dopings, in turn leading to a spatial dependence of $K_{FB}$ and $K_{FAB}$. Such inhomogeneities were observed in a scanning tunneling microscopy study of a sister blue bronze Rb$_{0.3}$MoO$_3$ \cite{Brun,Monceau2012}, where $K_F$ was found to vary across the surface from about $0.65b^*$ to about $0.80b^*$ ($b^*=2\pi/b$). The ARPES measurements on the 1D Mott-Hubbard insulator, SrCuO$_2$, considered the best evidence for spin-charge separation, also yield broad peaks much broader than the corresponding theoretical prediction (see Fig. 3 in \cite{Kim}). Hence, we believe that such broadening is an inherent feature of ARPES measurements on poorly conducting systems.

To account for these additional broadening effects, we convolve the lineshapes with a Gaussian of width $ = .07\pi b^{-1}$ (Fig. \ref{fig:MDC}, Fig. 5b and 5c). A considerable increase in broadening accounts very well for the experimental lineshapes, leading to a flat but finite MDC width at low energies while at higher energies the MDC widths grows as would be expected in a system with differing spin and charge velocities, Fig. 5d. Our theory thus accounts for the experimental MDC width at both low and high energies.

In conclusion, we have presented high resolution ARPES data for the CDW material K$_{0.3}$MoO$_3$.  Using non-perturbative field theoretic
techniques, we have argued that the features in the ARPES lineshapes can be understood primarily as arising from electronic correlations.  Narrowily drawn, this treatment has implications for the theory of cuprate ladders.  In particular, our prediction of a bound state signature will apply to the ARPES
response, already measured \cite{yoshida}, of such ladders.  More broadly drawn, this finding
suggests that electronic interactions may play a similarly important role in
other materials exhibiting CDW-like order, from the chalcogenides to the maganites to stripe ordered cuprates.

\begin{acknowledgements}
The authors are grateful for constructive conversations with Alan Tennant as well as useful comments by both
P. D. Johnson and by F.H.L. Essler. XJZ thanks financial support from the MOST of China (973 program No: 2011CB921703).
The work was also supported by the US DOE under contract number DE-AC02-98 CH 10886 (RMK, AMT, IZ). RMK and AMT also
thank the Galileo Galilei Institute for Theoretical Physics and the INFN for kind hospitality and support during the
completion of this work.
\end{acknowledgements}

\newpage

\onecolumngrid

\section{Supplementary material: Incoherent Soliton Excitations and Spin-Charge Separation in Blue Bronze}

\subsection{Determination of  Spin and Charge Velocities}

To determine the spin and charge velocities we begin with the experimental determination of the velocities
for $k_{||}>K_{FB}$, $v_R =0.47$eV$b\pi^{-1}$ and $k_{||}<K_{FB}$, $v_L=2.16$eV$b\pi^{-1}$.  (Here b is the
lattice spacing.)
For $k_{||}<K_{FB}$ we assume that the spectral response sees roughly equal
contributions from the the spinon and the holon (consistent with our theoretical analysis), and so
the measured velocity is a arithmetic mean of the spinon and holon velocities:
\begin{equation}
v_L = \frac{v_{sL} + v_{cL}}{2}.
\end{equation}
By how the MDC width grows with increasing binding energy (see Fig. 5d), we can estimate the difference of $v_{cL}$
and $v_{sL}$ :
\begin{equation}
v_{sL}-v_{cL} = 1.49 {\rm eV}b\pi^{-1}
\end{equation}
This gives the spin and charge velocities for $k_{||}<K_{FB}$ as
\begin{equation}
v_{sL} = 2.91 {\rm eV}b\pi^{-1}; ~~~ v_{cL} = 1.42 {\rm eV}b\pi^{-1}.
\end{equation}

For $k_{||}>K_{FB}$, we assume the spectral response is dominated by
the branch with the lesser velocity, again consistent with
our theoretical analysis, in this case the holon.  Thus it is this velocity that is directly measured:
\begin{equation}
v_{cR} = v_{R} = 0.47eVb\pi^{-1}.
\end{equation}
To determine the charge velocity $v_{cR}$ for $k_{||}>K_{FB}$ we resort to a theoretical analysis.
This gives the spin and charge velocities in terms of the (bare) velocities of the
bonding ($v_B$) and anti-bonding ($v_{AB}$) bands and the effective zero frequency component of the
Coulombic interaction, $V(0)$, as \cite{tsvelik_book}
\begin{eqnarray}
v_{cL} &=& v_{Av}(1+\frac{V(0)b}{2\pi v_{Av}})+\frac{\Delta v}{2};\cr\cr
v_{sL} &=& v_{Av}(1-\frac{V(0)b}{2\pi v_{Av}})+\frac{\Delta v}{2};\cr\cr
v_{cR} &=& v_{Av}(1+\frac{V(0)b}{2\pi v_{Av}})-\frac{\Delta v}{2};\cr\cr
v_{sR} &=& v_{Av}(1-\frac{V(0)b}{2\pi v_{Av}})-\frac{\Delta v}{2}.
\end{eqnarray}
where
\begin{eqnarray}
v_{Av} &=& \frac{v_{B}+v_{AB}}{2};\cr\cr
\Delta v &=& v_{B}-v_{AB}.
\end{eqnarray}
Using these relations we can then determine the remaining unknown velocities as well as $V(0)$:
\begin{equation}
v_{sR}=1.96{\rm eV}b\pi^{-1};~~v_{B}=2.16{\rm eV}b\pi^{-1};~~v_{AB}=1.21{\rm eV}b\pi^{-1};~~V(0)=-1.49eV.
\end{equation}
We see that our analysis gives a value of $V(0)$ consistent with blue
bronze being only moderately interacting.
It's negative value is consistent with (and necessary for) their being
a gap in the spin sector \cite{tsvelik_book}.

We can also show that the energy scale set by $V(0)$ governs the size of the gap of the spinons.  The spinon gap
is given by $\Delta \sim 0.1 \Delta_{1/2}$ \cite{esskon}
where $\Delta_{1/2}\sim W(g_{LR}/v_{Av})^{1/2}\exp(-\pi v_{Av}/g_{LR})$ \cite{review} is the gap
at half filling.  The prefactor $0.1$ relating the gap at 3/4-filling to the gap at 1/2-filling is a rough
estimate based on an $SO(6)$ symmetric theory.  Nonetheless we feel it is accurate to within a factor of 2.
While $g_{LR} \sim b V(K_{FAB}+K_{FB})$ involves the $K_{FAB}+K_{FB}$ Fourier component of the Coulomb interaction,
it is reasonable to suppose this component is of the same magnitude of $V(0)$.  This then gives $\Delta_{1/2} \sim W/2$
and so in turn $\Delta \sim W/20$.  Taking the bandwidth $W$ as $~1eV$, we see that it makes
a reasonable prediction for the magnitude of the observed spinon gap
of $\sim 60meV$.

\subsection{Description of  the Contribution of the Spectral Response from the Bound State}

To describe the contribution of the spectral response coming from the bound state we employ
a description of the ladders
appropriate where the Fermi velocities, $v_{AB}$ and $v_{B}$, of the ladders' two bands are equal.
We do so because then at low energies the low energy description of the doped ladders \cite{schulz,esskon}
admits a field theoretic
reduction governed by an SO(6) symmetry.  This field theory description of the ladder
allows us to compute analytically
the expected contribution to the spectral response of the boundstate.  While $v_{AB}$ different
than $v_{B}$ will distort
this response, we believe that this computation, appropriately modified for the differing velocities
(as described below),
will capture this response's gross features.

Following the procedure outlined in Ref. \cite{esskon} for the computation of response functions in
doped ladder systems,
we can estimate the contribution of the bound state to the single particle response to be
\begin{eqnarray}
G_{RB}(\omega,k) &\propto & \int d\gt_- \frac{g(\gt_-) e^{\gt_-/4}}{(\cosh(\gt_- )+\frac{1}{\sqrt{2}})^2}I(\omega,k,\gt_-);\cr\cr
I(\omega, k,\gt_-) &=& \int d\tau dx \frac{e^{i\omega \tau-ikx}}{(\tau-ix/v_c)^{1/4}}\int \frac{d\gt_+}{2\pi}
e^{-3\gt_+/4}e^{-2\Delta\sqrt{A(\gt_- )B(\gt_- )}|\tau|\cosh(\gt_+ )}
e^{-i2\Delta\sqrt{A(\gt_- )B(\gt_- )}{\rm sgn} (\tau) \frac{x}{v_s}\sinh(\gt_+ )};\cr\cr
g(\gt ) &=& \frac{e^{-5\gt/4}}{\cosh(\gt/2)};\cr\cr
A(\gt ) &=& \frac{e^{\gt/2}+\sqrt{2}e^{-\gt/2}}{2};\cr\cr
B(\gt ) &=& \frac{e^{-\gt/2}+\sqrt{2}e^{\gt/2}}{2}.
\end{eqnarray}
The above expression for the contribution of the bound state to the spectral function assumes
$v_B=v_{AB}$.  In order to then adapt it to our situation of $v_B\neq v_{AB}$, we use the following
heuristic.  For $k_{||}<K_{FB}$ we will take the spin and charge velocities, $v_s$ and $v_c$,
in the above to equal $v_{sL}$ and $v_{cL}$ (see previous section of the Supplementary Material for the definition
of these quantities).
For $k_{||}>K_{FB}$ we take $v_s$ and $v_c$ to instead
equal $v_{sR}$ and $v_{cR}$.  We still expect the velocities for the dispersion of the bound
state contribution to be different to the right and to the left of the Fermi point because these differing
velocities arise from a Gallilean transformation to a frame where $v_B=v_{AB}$.
What we do not expect is that the above captures correctly the intensity of the
boundstate contribution quantitatively (we might expect there
to be (large) corrections on the order of $\Delta v/v_B$), but merely provides a rough
qualitative guide.  However we can exclude the possibility that $v_{AB}$ and $v_B$ are so different
that the bound state does not
even exist because the experimental data suggests otherwise.

A final issue with which we must deal is the intensity of the bound state contribution relative to the
contribution of the spinon and holon alone.  We fix this intensity using the experimental intensities
as a rough guide.  Thus in Fig. 3c of the main text we take the peak intensity of the bound state
contribution to be
approximately 1/6 that of the peak intensity of the holon and spinon alone.

\subsection{Comparison of Theoretical and Experimental MDCs}

In this section we discuss further the theoretical and experimental
MDCs shown in Fig. 3 of the main text.  We already know from Fig. 5d
of the main text that the widths predicted by theory match those
measured. We now also see that over a range of binding energies
extending from 10meV to roughly 600meV that the theoretical MDCs match
approximately their measured counterparts. Beyond 600meV one begins to
see that the theoretical MDCs become asymmetrical and develop
structure within their central peak.  This is a consequence of the
bound state contribution to the MDCs, in particular the intensity.
Whereas the intensity of the spinon and holon contribution to the MDCs
falls off relatively rapidly with increasing binding energy, the
intensity of the computed bound state contribution does not.  This
contribution is found primarily at $k_{||}\sim E_{\rm binding~energy}/v_{cL}$ 
leading to a slight favoring of spectral weight towards this wave vector in the MDC. That this disagreement appears is not surprising given that our computation of the boundstate contribution to the spectral function is qualitative not quantitative.

\subsection{Temperature Dependency of Asymmetry in Dispersion}

\setlength\fboxsep{.5in}
\setlength\fboxrule{0pt}
\begin{figure}[t]
\subfigure{\includegraphics[width=0.45\textwidth,height=0.36\textwidth]{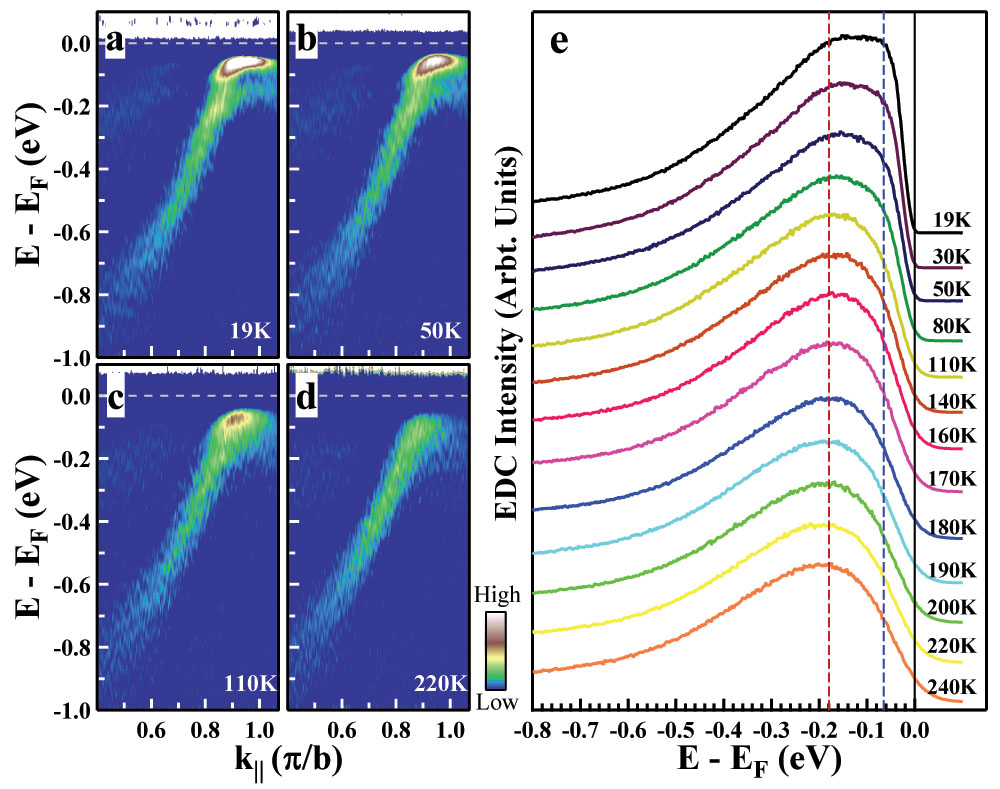}}
\subfigure{\fbox{\includegraphics[width=0.4\textwidth,height=0.36\textwidth]{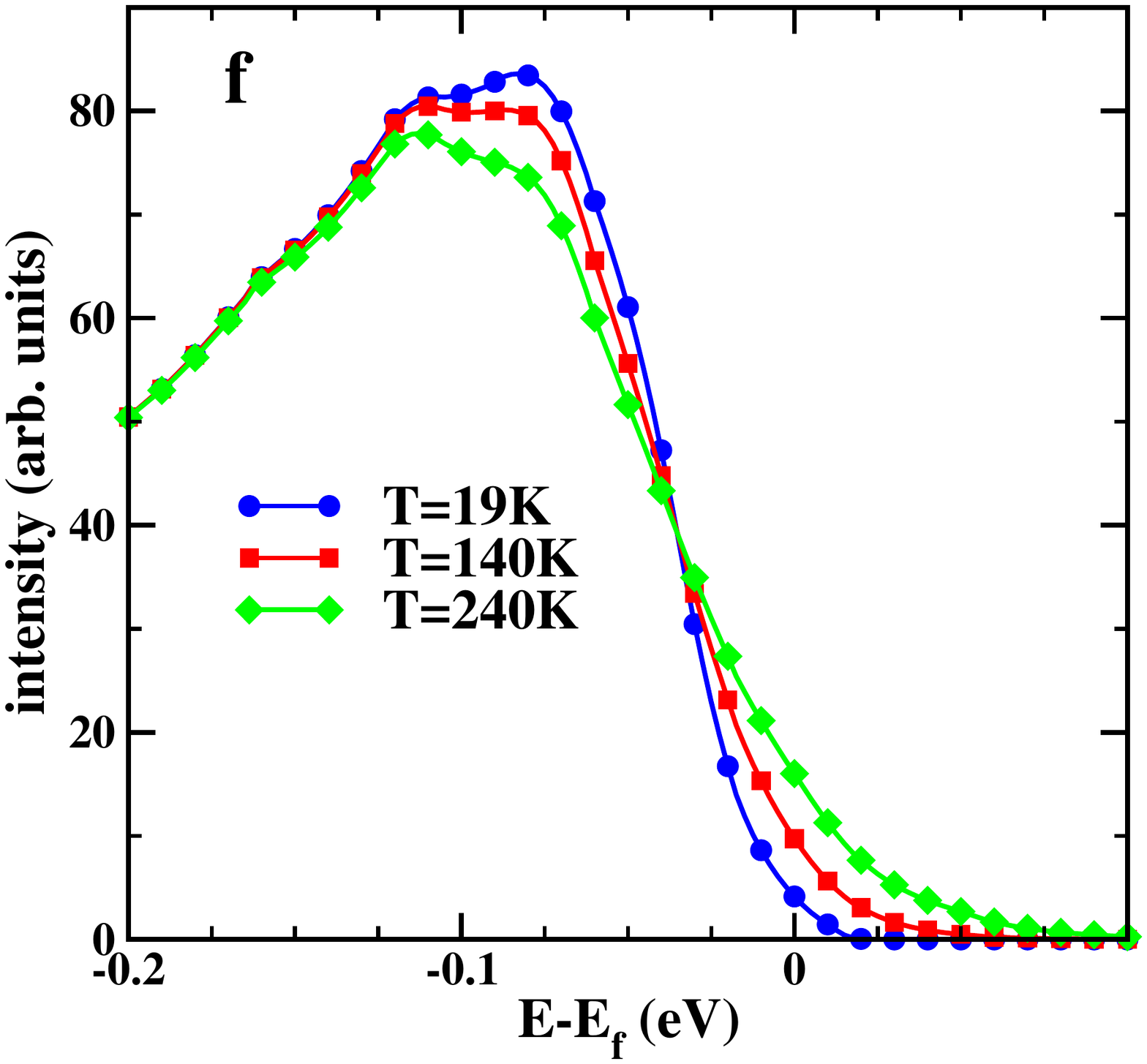}}}
\caption{
a) - d) EDC second derivative images of the B band along $\Gamma$-Y momentum cut at different temperatures.
The data are divided by the Fermi-Dirac function.
e) Extracted K$_{FB}$ EDCs at different temperatures. Two energy scales of incoherent soliton and bound state
are marked by blue and red dashed lines, respectively.
f) The theoretical $K_{Fb}$ EDC's for three different temperatures.
}
\label{fig:EDC}
\end{figure}

\begin{figure}[t]
\includegraphics[width=0.9\textwidth,height=0.4\textwidth]{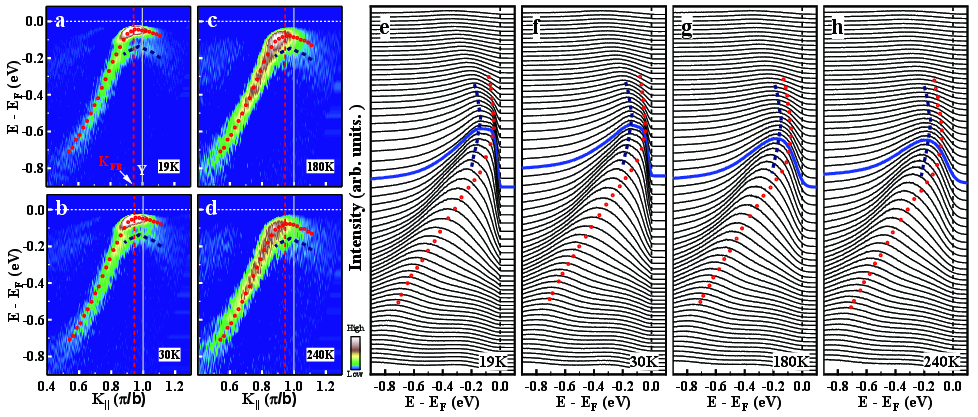}
\caption{a)-d) The second derivative of high resolution measurements of the band structure
along the $\Gamma-Y$ momentum cut with 1 meV energy resolution.  Measurements for four different
temperatures, below, at, and above the CDW transition temperature are shown.
e)-h) The corresponding EDCs. The red circles and blue dashes mark the dispersion of
the soliton/holon and the bound state/holon respectively.}
\label{fig:SM2}
\end{figure}

In this final section we present the temperature dependency of the asymmetry in the dispersion of spectral features to the left and to the right of $k=K_{FB}$.  We present in Figs. \ref{fig:EDC}, \ref{fig:SM2} the second derivative w.r.t. the energy of the photoemission spectra at four different temperatures, ranging from 19K to 240K, the final temperature being well above the CDW transition temperature, T$_c=180$ K.  We see the asymmetry in the dispersion is independent of temperature, persisting to well above $T_c$.  However we also see that as the temperature is increased spectral features are blurred.  The ability to distinguish between the contributions coming from the two branches (soliton/holon and bound state/holon) ceases at some temperature below T$_c$.  This is to be expected.  Sharp spectral features in correlated one dimensional systems see marked rounding even for temperatures a small fraction of the spectral gap \cite{finiteT}.

While it is a difficult task to compute the thermal broadening in a
strongly correlated electron system, the task is made easier if we
focus on the holon/spinon contribution to the spectral function.  We
can then follow the strategy of Ref. \cite{finiteT}.  At
least at low temperatures ($T \ll \Delta$), we can understand the spectral function's thermal broadening by
using the well understood form of the spectral contribution of the
gapless holon (that of a Luttinger liquid at finite temperature
\cite{Orgad}) while using the zero temperature form for the gapped
spinon.  While imperfect, this gives us a lower bound on the
broadening
of spectral lineshapes due to finite temperature.  The results can be
found in panel f of Fig. 1.  We see that the broadening due to
electronic correlations at finite temperature shares the same general
features
seen in the experiment (panel e of Fig. 1).  There is an increase in
spectral weight at energies, $E-E_{F} > 0 eV$ and a decrease in spectral
weight at energies around $E-E_{F} \sim -0.15eV$.  The increase in weight
at $E-E_{F} > 0 eV$ seen theoretically as temperature is increased
from 19K to 240K matches that seen experimentally.  On the other hand the decrease in weight at
$E-E_{F} \sim -0.15eV$ predicted theoretically by this calculation as
temperature is increased underestimates that seen experimentally.  While we again
emphasize that our calculation only provides a lower bound on thermal
broadening, it would certainly be reasonable (and expected) to need to ascribe
some broadening to the presence of phonons.

\begin{thebibliography}{99}
\bibitem{tsvelik_book} A. M. Tsvelik, {\it Quantum Field Theory in Condensed Matter Physics} (Cambridge University Press, Cambridge, 2003).
\bibitem{fabrizio} M. Fabrizio, Phys. Rev. B {\bf 48}, 15838 (1993).
\bibitem{fisher} L. Balents and M. P. A. Fisher, Phys. Rev. B{\bf 53}, 12133 (1996); H. H. Lin, L. Balents and M. P. A. Fisher,
Phys. Rev. B {\bf 58}, 1794 (1998).
\bibitem{schulz} H.J. Schulz, Phys. Rev. B{\bf 53}, R2959 (1996); ibid, cond-mat/9808167.
\bibitem{Khev} D. V. Khveshchenko and T. M. Rice, Phys. Rev. B {\bf 50}, 252
\bibitem{konik2} R. Konik, F. Lesage, A. W. W. Ludwig, and H. Saleur, Phys. Rev. B {\bf 61}, 4983 (2000);
R. Konik and A. W. W. Ludwig, Phys. Rev. B {\bf 64}, 155112 (2001);
R. Konik, F. Lesage, A. W. W. Ludwig, and H. Saleur, Phys. Rev. B {\bf 61}, 4983 (2000).
\bibitem{varma} C. M. Varma and A. Zawadowski, Phys. Rev. B {\bf 32}, 7399 (1985).
\bibitem{controzzi} D. Controzzi and A. M. Tsvelik, Phys. Rev. B {\bf 72}, 035110 (2005).
\bibitem{lee} H.C. Lee, P. Azaria, and E. Boulat, Phys. Rev. B {\bf 69}, 155109 (2004).
\bibitem{furusaki} M. Tsuchiizu and A. Furusaki, Phys. Rev. B {\bf 66}, 245106 (2002).
\bibitem{wu} C. Wu, W. V. Liu, and E. Fradkin, Phys. Rev. B {\bf 68}, 115104 (2003).
\bibitem{noack} R. M. Noack, S. R. White, and D. J. Scalapino, Phys. Rev. Lett. {\bf 73}, 882 (1994);
R. M. Noack, S. R. White, and D. J. Scalapino, Physica C 270, {\bf 281} (1996);
R. M. Noack, M. G. Zacher, H. Endres, and W. Hanke, cond-mat/9808020.
\bibitem{jeckelmann}  E. Jeckelmann, D. J. Scalapino, and S. R. White, Phys. Rev. B {\bf 58}, 9492 (1998).
\bibitem{weihong} Z. Weihong, J. Oitmaa, C. J. Hamer, and R. J. Bursill, J. Phys.:Condens. Matter {\bf 13}, 433 (2001).
\bibitem{poilblanc} D. Poilblanc, E. Orignac, S. R. White, and S. Capponi, Phys. Rev. B {\bf 69} , 220406R (2004).
\bibitem{2leg} A. M. Tsvelik, Phys. Rev. B{\bf 83}, 104405 (2011).
\bibitem{Greenblatt} M. Greenblatt, Chem. Rev. {\bf 91}, 965 (1988).
\bibitem{Monceau2012} P. Monceau, Advances in Physics {\bf 61}, 325 (2012).
\bibitem{Graham} J. Graham and A. D. Wadsley, Acta. Crystallogr.  {\bf 20}, 93 (1966).
\bibitem{Canadell} E. Canadell and M.-H. Whangbo, Chem. Rev. {\bf 91}, 965 (1991).
\bibitem{Mozos} J.-L. Mozos, P. Ordej\'{o}n, and E. Canadell, Phys. Rev. B {\bf 65}, 233105 (2002).
\bibitem{Johnston} D. C. Johnston, Phys. Rev. Lett. {\bf 52}, 2049 (1984).
\bibitem{Kwok} R. S. Kwok, G. Gruner and S. E. Brown, Phys. Rev. Lett. {\bf 65}, 365 (1990).
\bibitem{Pouget} J. P. Pouget  {\it et al.}, J. Phys. Lett. {\bf 44}, L113 (1983); Phys. Rev {\bf 43}, 8421 (1991).
\bibitem{Fleming1985} R. M. Fleming, L. F. Schneemeyer, D. E. Moncton, Phys. Rev. B {\bf 31}, 899 (1985). 
\bibitem{Sato} M. Sato {\it et al.}, J. Phys. C: Solid State Phys. {\bf 18}, 2603 (1985).
\bibitem{Fedorov} A. Fedorov {\it et.al.}, J. Phys.:Condens. Matter {\bf 12}, L191 (2000). 
\bibitem{Perfetti} L. Perfetti, S. Mitrovic, G. Margaritondo, M. Grioni, L. Forro, L. Degiorgi, and H. Hochst, 
Phys. Rev. B {\bf 66},  075107 (2002).
\bibitem{Ando} H. Ando {\it et.al.}, J. Phys.:Condens. Matter {\bf 17}, 4935 (2005).
\bibitem{Li} C. Li, {\it et al.}, J. Cryst. Growth, {\bf 285}, 81 (2005).
\bibitem{Liu} G. D. Liu {\it et.al.}, Rev. Sci. Instrum. {\bf 79}, 023105 (2008).
\bibitem{Essler2002} F. H. L. Essler and A. M. Tsvelik, Phys. Rev. B {\bf 65}, 115117 (2002).
\bibitem{Essler2005} F. H. L. Essler and A. M. Tsvelik, Phys. Rev. B {\bf 71}, 195116 (2005).
\bibitem{Konik} R. M. Konik, T. M. Rice and A. M. Tsvelik,  Phys. Rev. Lett. {\bf 96}, 086407 (2006).
\bibitem{Roscha} O. R\"{o}scha and O. Gunnarsson, Eur. Phys. J. B {\bf 43}, 11 (2005).
\bibitem{Brun} C. Brun {\it et.al.}, J. Phys.: Conf. Ser. {\bf 61}, 140 (2007).
\bibitem{finiteT} F. H. L. Essler and A. M. Tsvelik, Phys. Rev. Lett. {\bf 90}, 126401 (2003).  F. H. L. Essler and R. M. Konik,
Phys. Rev. B {\bf 78}, 100403; ibid, J. of Stat. Mech.: Theory and Experiment 2009, P09018 (2009).
\bibitem{supp} See supplemental material.
\bibitem{Kim} B. J. Kim {\it et. al.}, Nature Physics {\bf 2}, 397 (2006).
\bibitem{esskon} F. H. L. Essler and R. M. Konik, Phys. Rev. B {\bf 75}, 144403 (2007).
\bibitem{review} See Ch. 5 of F. H. L. Essler, and R. M. Konik, in {\it Ian Kogan Memorial Collection, From Fields to Strings:
Circumnavigating Theoretical Physics}, edited by M. Shifman, A. Vainshtein, and J. Wheater (World Scientific, Singapore, 2005);
cond-mat/0412421.
\bibitem{Hohennadler} M. Hohenadler, G. Wellein, A. R. Bishop, A. Alvermann, and H. Fehske, Phys. Rev. B {\bf 73}, 245120 (2006).
\bibitem{Ning} W.-Q. Ning, H. Zhao, C.-Q. Wu, and H.-Q. Lin, Phys. Rev. Lett. {\bf 96}, 156402 (2006).
\bibitem{yoshida} T. Yoshida, X. J. Zhou, Z. Hussain, Z.-X. Shen, A. Fujimori, H. Eisaki, and S. Uchida, Phys. Rev. B {\bf 80}, 052504 (2009).
\bibitem{Orgad} D. Orgad, Philos. Mag. B {\bf 81}, 377 (2001).
\end{thebibliography}
\end{document}